\documentclass[twocolumn,aps,prl,superscriptaddress,showpacs,secnumroman,showkeys]{revtex4-1}
\usepackage{amsmath,bm}%
\usepackage{graphicx}%
\usepackage{color}%

\begin{document}


\title{Matter-wave interference originates mass-angle correlation in fusion-fission}


\author{O.M. Gorbachenko}
\email[E-mail at:]{gorbachenko@univ.kiev.ua}
\affiliation{Faculty of Physics, Taras Shevchenko National University, Kyiv, 03680, Ukraine}
\author{S. Kun}
\email[E-mail at:]{ksy1956@gmail.com}
\affiliation{Canberra, Australia}



\begin{abstract}
Mass-angle correlation of fission fragments has been understood as manifestation of quasifission. We show that this is not so: the effect
can originate from correlation between fusion-fission amplitudes with different total spins signifying matter-wave interference in compound
nucleus processes. This resolves the well-known puzzle with the mass-angle correlation in the complete fusion sub-barrier reaction
$^{16}$O+$^{238}$U. Our finding is important for more reliable predictions of production cross sections for superheavy elements. Matter-wave
interference also produces quantum-classical transition to the time-orientation localization of the coherently rotating dinucleus in
quasifission.
\end{abstract}

\maketitle

In classical physics an individual binary collision is characterized by a single orbital momentum and, therefore, by a single impact parameter
corresponding to precisely defined transverse spatial position - spacial localization - of the incoming material particle. In contrast, in quantum
mechanics each single particle of the incident beam is represented by a coherent superposition
of partial waves with different orbital momenta.
Interference between the states with different
orbital angular momenta carried at once by the same incoming particle signifies matter-wave interference in the entrance channel.
Manifestation of matter-wave interference between spatially distinct
reaction paths - different impact parameters - in chemical reactions has been demonstrated
\cite{AoizZare18} by mapping the quantum collision problem onto the famous Young double-slit experiment \cite{Arndt14}.

We ask the question: Is matter-wave interference necessarily destroyed
 in the compound nucleus (CN) reactions? The modern understanding, based on the Bohr's independence hypothesis,
 assumes absence of the $S$-matrix total spin $(J)$ and parity ($\pi $) off-diagonal correlations \cite{Mitchell10}. This means that the CN formation produces effect of decoherence similar to that
 originated from detection of the which-path - which impact parameter - in the double-slit experiment \cite{Arndt14} thereby
destroying matter-wave interference and
leading to the fore-aft symmetry in CN reactions \cite{Mitchell10}.
However this conventional understanding is inconsistent with many data sets demonstrating a strong fore-aft asymmetry of the evaporation spectra.
The first well-known example of a strong fore-aft asymmetry of the evaporation spectra
was reported in \cite{Gugelot54} for the Pt(p,p') process.
Examples of strong fore-aft asymmetry of the evaporation proton spectra in nucleon-induced reactions with the targets $^{197}$Au, $^{208}$Pb,
$^{209}$Bi and $^{nat}$U are displayed in \cite{Benet08}.
Some representative data on the strong fore-aft asymmetry and anomalous, significantly stronger than $(1/\sin\theta )$, back-angle peaking of the proton and
$\alpha$-particle evaporation spectra demonstrating matter-wave interference in CN processes will be displayed in the extended version of this Letter.

Here we refer to the conclusive experimental evidence for the mass-angle correlation (MAC) for the sub-barrier reaction $^{16}$O+$^{238}$U
\cite{Hinde96}.
This MAC and anomalously large anisotropy of the fission fragments were interpreted \cite{Hinde96} as the
orientation-dependent quasifission \cite{ Hinde95}.
 The idea \cite{ Hinde95} was tested by measuring the evaporation residue cross section for the reaction $^{16}$O+$^{238}$U at the
 sub-barrier energies \cite{Nishio04}.
 The interpretation \cite{Hinde96}, \cite{ Hinde95} has been refuted by the experimental evidence that the complete fusion is the
 main process in the sub-barrier region while quasifission contribution is insignificant \cite{Nishio04}. Therefore MAC in the
 reaction $^{16}$O+$^{238}$U \cite{Hinde96} originates from matter-wave interference in fusion-fission and not in quasifission.

A correct understanding of competition between fusion-fission and quasifission
in reactions involving prolate deformed actinide nuclei
is important
for reliable predictions of production cross sections for superheavy elements \cite{Itkis15}, \cite{Hinde18}. For example, the insignificance of
quasifission, in particular, the orientation-dependent quasifission \cite{Hinde95}, in the complete fusion sub-barrier
reaction $^{26}$Mg+$^{248}$Cm led to the observation of the new superheavy nuclide $^{271}$Hs \cite{Dvorak08}.

In order to describe MAC
one can consider fission as a complex particle evaporation \cite{Ericson60}, \cite{Rossner84}. Then the result \cite{Kun94},\cite{Kun97a} for the CN $(J_1\neq J_2)$-correlation
is directly applicable to reproduce MAC and fore-aft asymmetry of fusion-fission angular distributions (FFAD).

   In this Letter we extend the more popular treatment (the transition-state statistical model (TSSM)) of the FFAD \cite{HalpStr58} by introducing
   the CN $(J_1\neq J_2 )$-correlation.
We represent FFAD in the form
$W(\theta)=\sum_{K}<|\sum_{Jc}F_{a,cK}^J(\theta,E)|^2>$.
Here
$F_{a,cK}^J(\theta,E)=(2J+1)S_{a,cK}^J(E)d_{0K}^J(\theta)$,
 $<...>$ stands for the energy averaging,
$d_{0K}^{J}(\theta)=D_{0K}^{J}(\alpha=0,\theta,\gamma=0)$,
 $D$-functions are the wave functions of the axially symmetric top,
 $a$ is index for the entrance channel,
 $K$ is projection of $J$ on the symmetry axis of the nucleus at the saddle-point and $c$ stands for the rest of indices of the fission channels.

Neglecting $J$-dependence of the potential phase shifts in the fission channels
we take $S$-matrix in the form $S_{a,cK}^{J}(E)=-i\exp(i\varphi_{a}^J)\delta S_{a,cK}^{J}(E)$, where $\varphi_{a}^J$ is the potential phase shift in the entrance channel while  $\delta S_{a,cK}^{J}(E)$ is given by the
pole expansion \cite{Bertsch18} with the residues $\gamma_\mu^{Ja}\gamma_\mu^{J,cK}$. Here, $\gamma$'s
are the real partial width amplitudes in the entrance and exit (fission) channels at the saddle-point with the $\mu$-index denoting the CN resonance level.
We assume that statistical properties of $\gamma$'s with fixed $J$-value are described by statistics of Gaussian Orthogonal Ensemble.
Note that, for $J\geq 1$,
inclusion of $K$-projections into the fission channel indices
  takes us to a regime of the weak coupling to the continuum,
 $\overline{(\gamma_\mu^{J,cK})^2}^\mu/D_J\approx 1/[2\pi(2J+1)]\ll 1$ ($D_J$ is the CN average level spacing). Therefore, the pole expansion \cite{Bertsch18}
  is an accurate approximation to the
$S$-matrix unitary representation in a regime of the strongly overlapping resonances \cite{GorinSel02}, $\Gamma /D_J\gg 1$,  with $\Gamma$
being the CN total decay width.
The $S$-matrix spin off-diagonal correlations result from the correlation between
$\gamma_{\mu_1}^{J_1 a}\sum_{c_1}\gamma_{\mu_1}^{J_1 ,c_1 K}$ and $\gamma_{\mu_2}^{J_2 a}\sum_{c_2}\gamma_{\mu_2}^{J_2 ,c_2 K}$ with $|J_1-J_2|\beta$ being
 the correlation length in the $(E_{\mu_1}^{J_1}- E_{\mu_2}^{J_2})$-space ($E_\mu^J$ are
the resonance energies) playing a role of quantum analogs of imaginary parts of the Ruelle-Pollicott resonances \cite{Ruelle86} (to be reported elsewhere).
In turn, this correlation
originates from the correlation between the CN resonance eigenstates $\phi_{\mu}^{J}$ of the type
\begin{eqnarray}
&&{ \int\int d{\bf R_1}d{\bf R_2} }
{ \left[ \overline{\phi_{\mu_1}^{J_1}({\bf r},{\bf R_1})\phi_{\mu_1}^{J_1}({\bf r},{\bf R_2}) \phi_{\mu_2}^{J_2}({\bf r},{\bf R_1})\phi_{\mu_2}^{J_2}({\bf r},{\bf R_2})}^{\bf r} \right. } \nonumber \\
&&{\left.- \overline{\phi_{\mu_1}^{J_1}({\bf r},{\bf R_1})\phi_{\mu_1}^{J_1}({\bf r},{\bf R_2})}^{\bf r} {~}\overline{\phi_{\mu_2}^{J_2}({\bf r},{\bf R_1})\phi_{\mu_2}^{J_2}({\bf r},{\bf R_2})}^{\bf r} \right].}
\label{WaveFunCorr}
\end{eqnarray}
Here ${\bf r}=({\bf r}_1,{\bf r}_2,...,{\bf r}_n)$ are coordinates of $n$
nucleons and  ${\bf R_{1,2}}$ are
 coordinates of the rest of the $(A-n)$ nucleons.
In Eq.~\eqref{WaveFunCorr} $\overline{(...)}^{\bf r}$ stand for the spacial averaging.
Note that upon the partial $(\mu_{1},\mu_{2})$-averaging (keeping $(E_{\mu_1}^{J_1}-E_{\mu_2}^{J_2})$  fixed with accuracy $\ll \beta$)
the main contribution to Eq.~\eqref{WaveFunCorr} is produced with
${\bf R_1}\to{\bf R_2}$.
Seemingly paradoxically, the problem of the cross-symmetry correlation relates to
the Wigner dream \cite{Gard72} to develop a theory of correlations between reduced widths within
 a single symmetry sector (to be reported elsewhere).

The final result, for the fission fragments sufficiently lighter than the average fragment mass, reads
\begin{eqnarray}
&&W(\theta)\propto\sum_{J_1,J_2} (2J_1+1)(2J_2+1) (T_a^{J_1}T_a^{J_2})^{1/2} \nonumber \\
&&\exp(i\varphi_{a}^{J_1}- i\varphi_{a}^{J_2})[1+
(\beta /\Gamma )|J_1-J_2|]^{-1} \nonumber \\
&&\sum_{K=-{\rm min}(J_1,J_2)}^{{\rm min}(J_1,J_2) } [B_{J_1}(K)B_{J_2}(K)]^{1/2}d_{0K}^{J_1}(\theta) d_{0K}^{J_2}(\theta)
\label{Wcorrect}
\end{eqnarray}
with $B_J(K >J)=0$, $B_J(K\leq J)=p_K/\sum_{{\tilde K}=-J}^{J}p_{{\tilde K}}$, $p_K=\exp[-K^2/2K_0^2]$.
Here, $K_0^2={\cal J}_{eff} T_b/\hbar^2$, ${\cal J}_{eff}^{-1}={\cal J}_{perp}^{-1}-{\cal J}_{par}^{-1}$,
${\cal J}_{par}$ and ${\cal J}_{perp}$ are the nuclear moments of inertia for rotations around the symmetry axis
and a perpendicular axis, $ T_b=[8(E-B_f-E_{rot})/A]^{1/2}$ is the nuclear temperature
at the saddle-point, $B_f$ is the fission barrier, $E$ is the excitation energy and $E_{rot}$ is the
rotational energy.
For $\beta /\Gamma \gg 1$ the $(J_1\neq J_2 )$-correlations decay much
faster than the CN average life-time thereby leading to the TSSM result \cite{HalpStr58}. The time power spectrum, $P(\theta ,t)$, is given by Eq.~\eqref{Wcorrect} with
$\exp[-(\Gamma +\beta |J_1-J_2|)t/\hbar]$ instead of $[1+(\beta /\Gamma )|J_1-J_2|]^{-1}$. The time-dependent FFAD is
$\propto \int_0^td\tau P(\theta ,\tau )$.

Eq.~\eqref{Wcorrect} reproduces fore-aft asymmetry and MAC for the mass-asymmetric fission due to the
$(J_1\neq J_2 )$-contributions with odd values of
$(J_1+J_2)$, {\sl i.e.}, $\pi_1\neq\pi_2$. The fore-aft symmetry is restored destroying MAC
for $\beta_{\pi_1\neq\pi_2}\gg\Gamma$ leading to
 $W(\theta)\to W(\theta)+W(\pi-\theta)$, like for the mass-summed FFAD.
 However, such a fore-aft symmetry does not necessarily mean a complete mass-symmetrization but can also imply that the
 mass-asymmetric deformed system at the saddle-point is oriented in opposite directions with equal probabilities.
 Yet, weakening of MAC
 with a decrease of the mass-asymmetry can be reproduced by increasing   $\beta_{\pi_1\neq\pi_2}/\Gamma$
 with $\beta_{\pi_1\neq\pi_2}/\Gamma\to\infty$ for the mass-symmetric fission.

On the initial stage the colliding ions form a dinuclear system. A distribution of its orientations, $P_{K=0}(\theta )$, is identified with the
time power spectrum at $t=0$ for dissipative heavy-ion collisions \cite{Kun97b} or
for heavy-ion scattering
\cite{Kun01}.
It follows that
$P_{K=0}(\theta )$ is formally given by Eq.~\eqref{Wcorrect} with $\beta=0$ and $K_0^2=0$ ($B_J(K)=\delta_{0K}$).
We expand
$\varphi_{a}^{J}=\Phi (J-{\bar J})+{\dot \Phi}(J-{\bar J})^2/2$, where ${\bar J}=\sum_{J=0}^\infty JT_a^J/\sum_{J=0}^\infty T_a^J $ while $\Phi$ and $\dot{\Phi}$
have a meaning of the classical deflection angle and its derivative in the entrance channel.
$P_{K=0}(\theta )$ is peaked at $\theta\simeq \Phi$ having width
$\Delta_{K=0}\simeq [(1/<J^2>)+{\dot \Phi}^2<J^2>]^{1/2}$, where $<J^2>=\sum_{J=0}^\infty (2J+1)J^2T_a^J/\sum_{J=0}^\infty (2J+1)T_a^J $.
Then, for
$\Delta_{K=0}\leq 1$, we are dealing with a fusion of the dinuclear system preferentially orientated along the $\Phi$-direction.
It follows from Eq.~\eqref{Wcorrect} that the pre-fusion phase relations (initial conditions) in the entrance channel, $(\varphi_{a}^{J_1}- \varphi_{a}^{J_2})$, are
not forgotten even for the complete $K$-equilibration, $K_0^2\gg <J^2>$, providing  $\beta /\Gamma $ is a finite quantity.
 This means that if the same CN is formed but with different phase relations in
 the entrance channel (different collision partners) the FFADs will be different for a finite value of $\beta /\Gamma $.
For $\Delta_{K=0} \geq \pi$, the ($J_1\neq J_2$)-correlation and MAC
 are strongly suppressed already on the pre-fusion stage
even for a finite value of $\beta /\Gamma $.

The  sub-barrier $^{16}$O+$^{238}$U fusion-fission \cite{Nishio04} demonstrates both MAC and anomalous anisotropy
\cite{Hinde96}.
To illustrate the effect of the orientation-dependent fusion-fission
we focuse on the energy interval  $E_{c.m.}=72.8-75.6$ MeV ($E=34.5-37.3$ MeV, $\Gamma /D_J\simeq 10^{25}$) for which the interpretation \cite{Hinde96}, \cite{Hinde95}
suggested absence of the fusion-fission.
Employing a standard statistical model we find
 that to reproduce the experimental value of $\sigma_{ER}({\rm 4n})/\sigma_{fiss}\approx 6\times 10^{-5}$ for $E_{c.m.}=72$ MeV \cite{Nishio04} it is
necessary to take $B_f=4.15$ MeV instead of
$B_f=1.5$ MeV \cite{Sierk86}.
Here $\sigma_{fiss}$ and $\sigma_{ER}({\rm 4n})$ are the fission and
4n evaporation residue cross sections.
For $B_f=1.5$ MeV, a statistical model predicts  $\sigma_{ER}({\rm 4n})/\sigma_{fiss}\approx 10^{-11}$ for $E_{c.m.}=72$ MeV
instead of the experimental value of $\approx 6\times 10^{-5}$ \cite{Nishio04}.
For $T_b\approx 1$ MeV and
$B_f=4.15$ MeV, the pre-equilibrium fission \cite{Ramamurthy85},\cite{Vorkapic95},\cite{Liu96}, which anyway can not reproduce the fore-aft asymmetry
and the associated MAC,
is insignificant.

With a decrease of $E_{cm}$ from 75.6 MeV to 72.8 MeV $(A_{exp}-1)$ increases from $\approx 0.85$ to $\approx 1.6$ \cite{Hinde96} ($A=W(\pi)/W(\pi/2)$).
In contrast, the TSSM  predicts a decrease of $(A_{TSSM}-1)$ from $\approx 0.16$ to $\approx 0.1$ \cite{Hinde96}, {\sl i.e.},
 $<J^2(75.6 {\rm MeV})>/<J^2(72.8 {\rm MeV})>\approx 1.6$,
With $K_0^2= 200$, $<J^2(72.8 {\rm MeV})>=80$ and $<J^2(75.6 {\rm MeV})>=128$.
Then, to reproduce $A_{exp}$ within the TSSM one has to take $K_0^2(75.6 {\rm MeV})\approx 37$ and
$K_0^2(72.8 {\rm MeV})\approx 12$ instead of $K_0^2\approx 200$ with ${\cal J}_{eff}$ from \cite{Sierk86}.

In our interpretation we take into account the orientation dependence of the sub-barrier fusion for the target having a prolate deformation \cite{Stokstad81}.
Then
we deal with a fusion of the strongly mass asymmetric dinuclear system with $K=0$ which is preferably oriented parallel to the beam axis with its
projectile-like tip
pointing opposite to the beam direction.
Therefore we take $\Phi=\pi$ for both $E_{cm}$=72.8 MeV and 75.6 MeV.

To minimize a width of $P_{K=0}(\theta )$ for the smaller sub-barrier energy $E_{c.m.}=72.8$ MeV we take
  ${\dot \Phi}=0$.
Then, for $T_a^J\propto \exp[-J^2/(<J^2>]$ with  $<J^2(72.8 {\rm MeV})>=80$,
$P_{K=0}(\theta )$ is peaked at $\theta=\pi$ having a width of
$\Delta_{K=0}$ $\approx 6^\circ$.
Admixture of the states with $K=1,2$ will increase dispersion of the pre-fusion dinuclear orientation.

Since $<J^2>/(4K_0^2)\approx 0.1-0.16\ll 1$ we take $K_0^2=\infty$  ($B_{J}(K)=1/(2J+1)$) resulting in isotropic FFAD in the absence
of the $(J_1\neq J_2 )$-correlation.
 Now the mass-summed FFAD, obtained by symmetrizing Eq.~\eqref{Wcorrect} about $\theta=\pi/2$, depends on a single parameter $\beta/\Gamma$.
 In Fig. 1, $A=2.6$ is reproduced with $\beta/\Gamma=1.66$.

 For $<J^2(75.6 {\rm MeV})>$=128 and this same value of $\beta/\Gamma=1.66$,
   $A_{exp}(75.6 {\rm MeV}) =1.85$  is reproduced in Fig. 1 with
  $|{\dot \Phi}|=3.6^\circ $ corresponding to a width of $P_{K=0}(\theta )$ of $\approx 8^\circ$ with its tail extended up to $\approx 135^\circ$.
   This demonstrates that a fast growing of $(A-1)$ with the energy decrease
  is due to the moderate reduction
  of a dispersion of the pre-fusion dinuclear orientation with a decrease of $E_{c.m.}$.
\begin{figure}[htbp]
    \includegraphics[width=0.48\columnwidth]{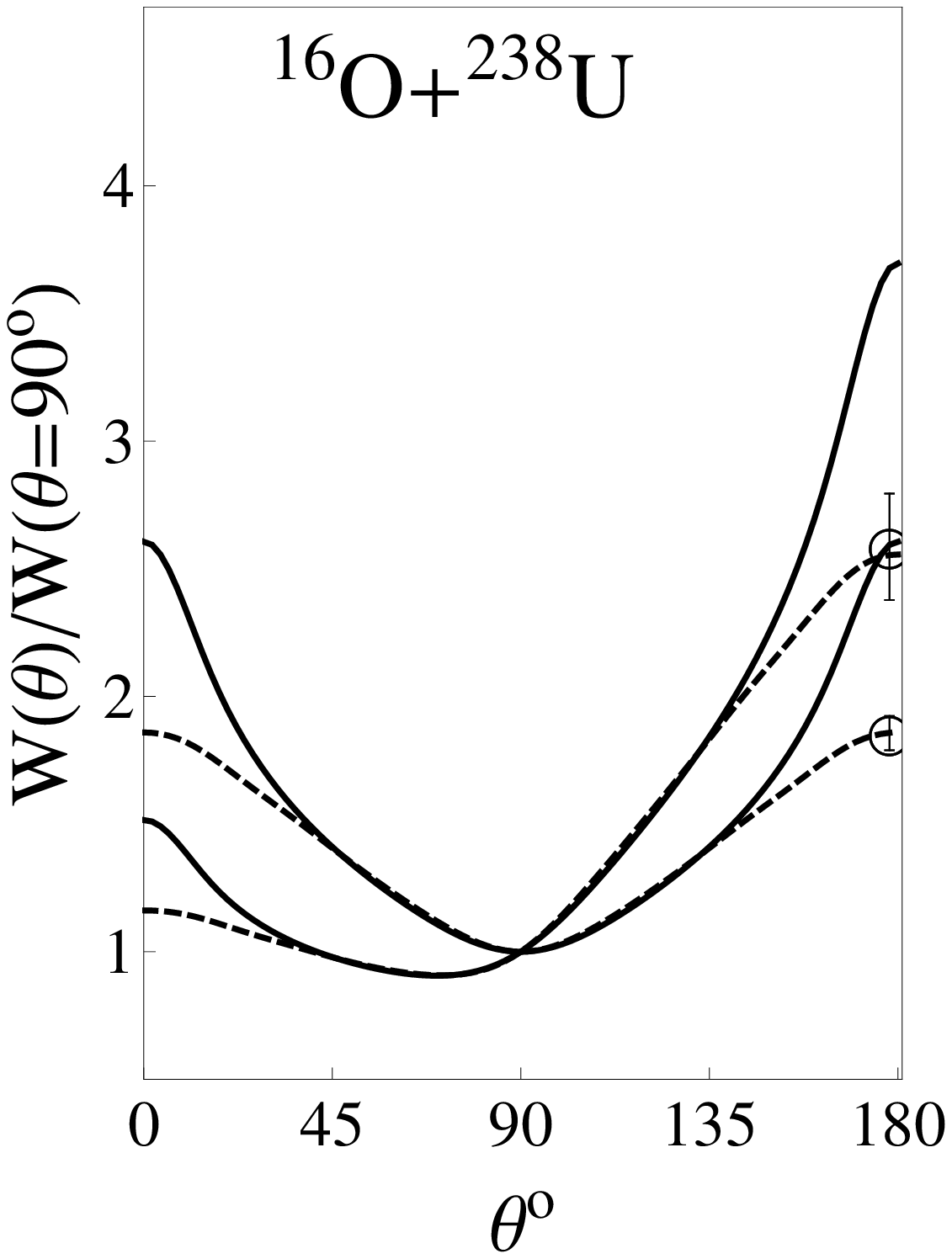}
    \includegraphics[width=0.48\columnwidth]{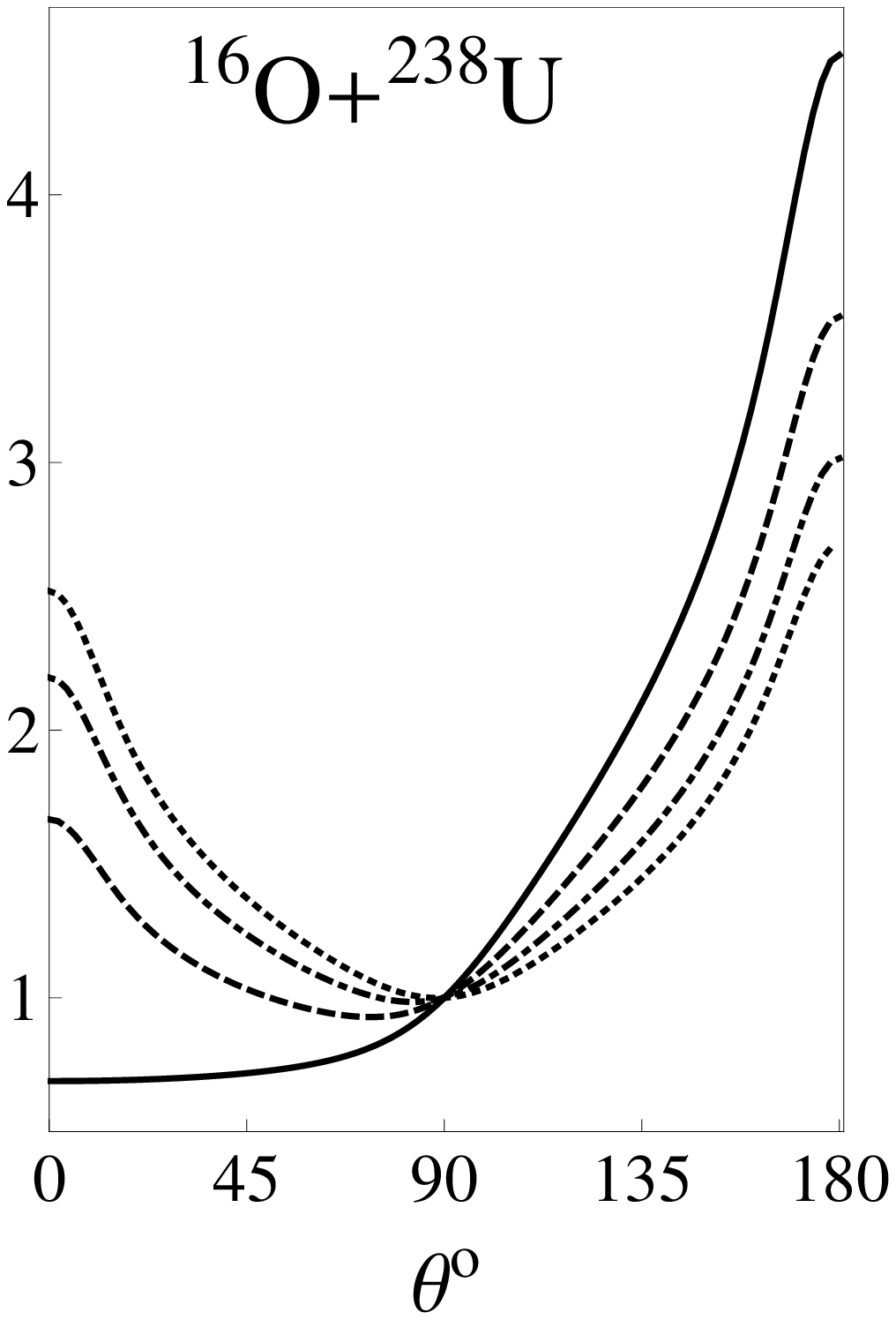}
	\caption{Left panel: Solid and dashed curves are FFADs calculated with $K_0^2=\infty$ for $^{16}$O+$^{238}$U collision at
 $E_{c.m.}=72.8$ MeV and 75.6 MeV, respectively. Symmetric and asymmetric about 90$^\circ$ curves correspond to light fragments and mass-summed FFADs,
  respectively (see text). Dots with error bars are experimental
 anisotropies for mass-summed FFADs \cite{Hinde96}.
Right panel: Light fragment FFADs calculated for $^{16}$O+$^{238}$U collision at
 $E_{c.m.}=72.8$ MeV  with $K_0^2=\infty$. Solid, dashed, dashed-doted and dotted curves correspond to
 $\beta_{\pi_1\neq\pi_2}/\Gamma=1.66$, 4, 10 and 50, respectively (see text).}
	\label{fig_1}
\end{figure}
With
 $\Gamma\approx \Gamma_f\approx T_b\exp(-B_f/T_b)/(2\pi )\approx 2.5$ keV ($T_b=1$ MeV, $B_f=4.15$ MeV) we obtain $\beta\approx 4$ keV and
 $\hbar /\beta\approx 1.65\times 10^{-19}$ sec.
 Therefore, the characteristic time for a decay of the CN  $(J_1\neq J_2 )$-correlation
 is about 3 orders of magnitude longer than the CN thermalization time, $\hbar/\Gamma_{spr}\simeq 1.3\times 10^{-22}$ sec, where $\Gamma_{spr}\simeq 5$ MeV is a width of the giant resonances.

MAC has been revealed in \cite{Hinde96} from asymmetry of the fragment mass distribution at the back-angles.
For the lowest energy reported, $E_{c.m.}=72.8$ MeV, it has been found that
 the light fragment yield, $M_L\leq 100$, integrated over the angular
range $110^\circ\leq \theta_{c.m.}\leq 155^\circ $ exceeds that for the heavy fragments, $M_H\leq 150$,
by a factor of 1.8$\pm 03$. To reproduce this experimental value we need $r=R_1/R_3=1.8$, where $R_{1(3)}=\int_{110^\circ (25^\circ)}^{155^\circ(70^\circ)}d\theta\sin\theta W(\theta )$. We calculate $r$ with $K_0^2=\infty$ and the previously fixed $\Phi=\pi$, ${\dot \Phi}=0$, $<J^2>=80$
 first with $\beta_{\pi_1=\pi_2}/\Gamma=\beta_{\pi_1\neq\pi_2}/\Gamma=1.66$. We obtain $r=2.7$ instead of the experimental value 1.8.
To reduce the calculated value of $r$ we increase $\beta_{\pi_1\neq\pi_2}/\Gamma$ but keep $\beta_{\pi_1=\pi_2}/\Gamma=1.66$ unchanged (to have the unchanged  mass-summed FFAD). The value $r=1.8$ is obtained with  $\beta_{\pi_1\neq\pi_2}/\Gamma=3.36$. The corresponding FFADs calculated with
$\beta_{\pi_1\neq\pi_2}/\Gamma=3.36$ for both the energies $E_{c.m.}=72.8$ MeV and 75.6 MeV are displayed in Fig. 1.
FFADs for the heavy fragments, $M_H\geq 150$, can be obtained by the reflection about $\pi/2$, $W_H(\theta)=W_L(\pi -\theta )$,
  reproducing the significant MAC \cite{Hinde96}.
Weakening
of the fore-aft asymmetry and MAC for $E_{c.m.}=72.8$ MeV with increase of $\beta_{\pi_1\neq\pi_2}/\Gamma$ is demonstrated in Fig. 1.
This weakening also illustrates evolution of FFADs for a transition from mass-asymmetric to mass-symmetric fission.

 While the analysis \cite{Hinde96} produced a conclusive evidence for MAC the mass-angle distribution is
 even more sensitive, informative and visually transparent experimental observable to demonstrate the effect.
 For example, MAC is unnoticeable in
the fragment mass distribution for the $^{19}$F+$^{238}$U reaction \cite{Pullanhiotan17}. Yet MAC is clearly seen in the mass-angle distribution,
Fig. 4 in \cite{Pullanhiotan17}.

 Suppose that, on the pre-fusion stage, the prolate shape of the target
 nucleus becomes unstable changing to the oblate shape with the symmetry axis oriented along and/or perpendicular to the beam direction.
 One may imagine a coexistence of the three pre-fusion configurations coherently
  contributing
 into a formation of the same CN. Can interference between the three fusion-fission amplitudes produce
oscillations in the FFAD?

 In case of the noticeable suppression of the evaporation residue yield indicating
 quasifission
 we would be led to deal with the  $(J_1\neq J_2 )$-correlation
 at the conditional saddle-point.
 The corresponding expression for the angular distribution can be envisaged from the
consideration of the angular distributions in dissipative heavy-ion collisions \cite{Kun97b}. The result for the
 time power spectrum is given by Eq.~\eqref{Wcorrect} with
$\exp(-\Gamma t/\hbar)\exp[-i\omega t(J_1-J_2)-\beta |J_1-J_2|t/\hbar]$ instead of $[1+(\beta /\Gamma )|J_1-J_2|]^{-1}$,
where $\omega$ is a real part of the angular velocity of the coherent rotation of the dinuclear system.
In this resulting expression
 possible time-dependencies of
 $\omega$ and  $K_0^2$ can be taken into account \cite{Dossing85}.
The time power spectrum for quasifission describes  the classical-like rotation of the deformed system having
a classically single $J$-value \cite{Kun97b}.
Yet, this classical-mechanics picture originates from a coherent superposition of the dinuclear states with
different quantum-mechanical $J$-values.
Determination of the impact parameter of the incident particle using a which-path detector destroys the coherent superpositions
 \cite{AoizZare18}, \cite{Arndt14} of the dinuclear states with different quantum-mechanical $J$-values
 disabling the
 quantum-classical transition to the macroscopic-like time-orientation localization and the classical-like rotation.
Therefore it is matter-wave interference that produces quantum-classical transition to
the time-orientation localization/correlation for the macroscopic-like rotation of the deformed system.
This interpretation
is complementary to the traditional understanding that the emergence of macro-world described
by classical physics originates from decoherence destroying matter-wave interference \cite{Arndt14}.
  For $\beta \ll\Gamma$ and $\Phi\approx 0$ the fore-aft asymmetry factor for quasifission fragment angular distribution
is $\propto\cosh [(\pi -\theta)\Gamma /(\hbar\omega )] $ demonstrating yet again that matter-wave
  interference produces a classical-like
 rotation of the osculating complex with a classically single $J$-value (fixed $\omega$),
 Fig. 14 in \cite{Herschbach86}.

 The wave nature of the heated organic macromolecules has been firmly established, in a model-independent way, raising a question of matter-wave interference
for biologically functioning entities of elevated temperature carrying the code of self-replication such as viruses and bacteria \cite{Geyer16}.
In this Letter we have proposed that matter-wave interference
plays an important role in nuclear fission.
 We have demonstrated that, contrary to the conventional understanding,
 MAC does not necessarily indicate hindrance of the CN formation due to
  quasifission but can originate from the CN spin off-diagonal phase correlation signifying matter-wave interference in fusion-fission.
  One of the central conventionally counter-intuitive problems of the proposed interpretation, which has not been explicitly addressed in the study of matter-wave interference with complex molecules \cite{Arndt14},\cite{Geyer16},
  is to justify
  the anomalously slow cross-symmetry phase relaxation
in classically chaotic many-body systems as compared to the relatively fast phase and energy relaxation (thermalization) within a single symmetry sector.

\begin{acknowledgments}

\end{acknowledgments}


\end{document}